%% file: unletter.tex
\documentclass[english,american,notitlepage,usenames,dvipsnames,svgnames,table]{revtex4-1}
\usepackage[T1]{fontenc}
\usepackage[latin9]{inputenc}
\usepackage{float}
\usepackage{amsmath}
\usepackage{esint}

\makeatletter
\usepackage{graphicx}

\usepackage{cfr-lm}
\usepackage{microtype}

\usepackage[american]{babel}

\newcommand{\Lagr}{\mathcal{L}}

\let\Im\relax
\DeclareMathOperator{\Im}{Im}

\DeclareMathOperator{\Tr}{Tr}

\DeclareMathOperator{\sgn}{sgn}

\input{preamble-notikz.tex}

\makeatother

\usepackage{babel}
\begin{document}

\title{Schwinger Pair Production at Finite Temperature}

\author{Leandro Medina and Michael C.~Ogilvie}

\address{Dept.~of Physics, Washington University, St.~Louis, MO 63130 USA}

\email{mco@wustl.edu, leandro@wustl.edu}

\begin{abstract}
Thermal corrections to Schwinger pair production are potentially important
in particle physics, nuclear physics and cosmology. However, the lowest-order
contribution, arising at one loop, has proved difficult to calculate
unambiguously. We show that this thermal correction may be calculated
for charged scalars using the worldline formalism, where each term
in the decay rate is associated with a worldline instanton. We calculate
all finite-temperature worldline instantons, their actions and fluctuations
prefactors, thus determining the complete one-loop decay rate at finite
temperature. The thermal contribution to the decay rate becomes nonzero
at a threshold temperature $T=eE/2m$, above which it dominates the
zero temperature result. This is the lowest of an infinite set of thresholds
at $T=neE/2m$. The decay rate is singular at each threshold
as a consequence of the failure of the quadratic approximation to the worldline
path integral. We argue that that higher-order effects  will make the decay rates finite
everywhere, and model those effects by the inclusion of  hard thermal loop
damping rates. We also demonstrate that the formalism developed here
generalizes to the case of finite-temperature pair production in
inhomogeneous fields.
\end{abstract}
\maketitle

\section{Introduction}

Pair production in an external field is a form of semiclassical tunneling
and has applications in many areas of physics~\cite{Parker:1969au,Zeldovich:1971mw,Hawking:1974rv,Casher:1978wy,Ringwald:2001cp,Kharzeev:2005iz}.
Here we present a complete first-principles calculation of the one-loop
thermal correction to the pair production rate of charged scalars
in a static electric field. The effect of pair production in a background
electric field at zero temperatue was derived first by Euler and Heisenberg~\cite{Heisenberg:1935qt}
and subsequently rederived by Schwinger~\cite{Schwinger:1951nm}
using modern field-theoretic techniques. % proposedThe physics of
the pair production is simple: when the energy contained in an external
electric field is large enough, it becomes energetically favorable
to produce charged pairs which screen the external field. From a modern
perspective, the presence of an external electric field over a large
spatial region creates a metastable state, which decays by the nucleation
of charged-particle pairs. In this way, it is similar to the false
vacuum decay~\cite{Langer:1967ax,Coleman:1977py,Callan:1977pt}.
This similarity is most clearly seen in the worldline formalism, as
shown in the calculation of the zero-temperature pair production rate
by Affleck \emph{et al.}~\cite{Affleck:1981bma}. The inclusion of
thermal effects naturally increases the rate at which the metastable
state decays~\cite{Affleck:1980ac}.

Schwinger's expression for the decay rate is obtained from the imaginary
part of the one-loop effective action of charged particles in a constant
external electric field. For charged scalars, the one-loop zero-temperature
decay rate is
\begin{equation}
\Gamma=\frac{\left(eE\right)^{2}}{\left(2\pi\right)^{3}}\sum_{p=1}^{\infty}\frac{\left(-1\right)^{p+1}}{p^{2}}\exp\left[-\frac{m^{2}}{eE}\pi p\right]\label{eq:Schwinger-scalar-rate}
\end{equation}
with a similar result for fermions. The factor of $1/e$ in the exponent
signals that this is a nonperturbative result. These results have
been extended in a number of ways~\cite{Brezin:1970xf,Popov:1973az,Marinov:1977gq,Dunne:2004nc,Gies:2005bz,Dunne:2005sx,Dunne:2006st,Dunne:2012vv}.
One obvious extension is to to nonzero temperature and density. In
the case of external \emph{magnetic} fields, the properties of the
thermal one-loop effective action are well-known~\cite{Dittrich:1979ux,Elmfors:1993wj}.
However, in the case of electric fields, there has been no clear consensus
on the form or even the existence of one-loop thermal corrections
to the zero-temperature decay rate~\cite{Loewe:1991mn,Elmfors:1994fw,Hallin:1994ad,Ganguly:1995mi,Ganguly:1998ys,Gies:1998vt,Gies:1999vb}.
Although the formal expression for the decay rate can be readily constructed
using, say, Schwinger's proper time formulation, the analytic structure
of the resulting formulae is quite intricate and leads to structural
ambiguities~\cite{Gies:1998vt}. It has been suggested that the one-loop
thermal contribution to the decay rate may be zero, but there is no
obvious symmetry principle that would lead to this conclusion. The
worldline formalism has proven to be a very powerful tool in quantum
field theory at zero temperature, capable of reproducing and extending
Schwinger's result~\cite{Affleck:1981bma,Dunne:2005sx,Dunne:2006st}
as well as providing a compact, powerful framework for the calculation
of gauge theory amplitudes~\cite{Bern:1991aq,Strassler:1992zr} We
will show that the worldine formalism can be used to calculate the
thermal corrections to Schwinger's one-loop result. To the best of
our knowledge, this is the first time the worldline formalism has
been used to calculate a nonperturbative finite-temperature effect.
%We will show the worldline formalism~\cite{Affleck:1981bma,Dunne:2005sx,Dunne:2006st}%is capable of resolving these ambiguities.We
We restrict ourselves here to the simplest case of charged scalars in
QED, and will return to the case of fermions in QED and QCD in later
work. The extension of the worldline formalism to fermions presents
no difficulty~\cite{Schubert:2001he,Dunne:2005sx}. The case of QCD
is relevant, for instance, in phenomenological flux-tube models of
quark-antiquark pair production during hadronization in heavy ion
collisions~\cite{Casher:1978wy,Andersson:1983ia,Bass:1998ca}.

\section{The $T=0$ case in the worldline formalism}

We begin by reviewing and extending the work of Affleck \emph{et al.}~\cite{Affleck:1981bma}
for the $T=0$ case of a scalar field. The Lagrangian is given by
\begin{equation}
\Lagr=\left(D_{\mu}\phi\right)^{*}\left(D_{\mu}\phi\right)+m^{2}\phi^{*}\phi
\end{equation}
with a covariant derivative $D_{\mu}\equiv\partial_{\mu}+ieA_{\mu}$
where $A_{\mu}$ provides the constant background electric field in
Euclidean space. The partition function $Z[A]$ and effective action
$W[A]$ are functionals of the background field:
\begin{equation}
Z\left[A\right]=e^{-W[A]}=\int\left[d\phi\right]\left[d\phi^{*}\right]e^{-\int d^{4}x\Lagr}.
\end{equation}
The integral over the scalar fields can be done exactly, yielding
a functional determinant which can be written as an integral over
proper time:
\begin{equation}
W\left[A\right]=-\int_{0}^{\infty}\frac{ds}{s}\Tr\left[e^{-s\left(-D^{2}+m^{2}\right)}\right]=-\int_{0}^{\infty}\frac{ds}{s}e^{-sm^{2}}\left\langle x\left|e^{-s\left(-D^{2}\right)}\right|x\right\rangle .
\end{equation}
In the worldline formalism, the trace is given as a path integral
over closed worldline paths $x_{\mu}(\tau)$
\begin{equation}
W\left[A\right]=-\int_{0}^{\infty}\frac{ds}{s}e^{-sm^{2}}\oint\left[dx_{\mu}\right]\exp\left\{ -\int_{0}^{s}d\tau\left[\frac{1}{4}\left(\frac{dx_{\mu}}{d\tau}\right)^{2}+ieA_{\mu}\frac{dx_{\mu}}{d\tau}\right]\right\}
\end{equation}
with boundary condition $x_{\mu}\left(0\right)=x_{\mu}\left(s\right)$.
We perform first a saddle point approximation to the integral over
$s$ and then find instanton solutions to the equations for $x_{\mu}(\tau)$.
By rescaling $s$ and $\tau$, we can put $W[A]$ into the form~\cite{Dunne:2005sx}
\begin{equation}
W\left[A\right]=-\int_{0}^{\infty}\frac{ds}{s}e^{-s}\oint\left[dx_{\mu}\right]\exp\left\{ -\int_{0}^{1}du\left[\frac{m^{2}}{4s}\dot{x}^{2}+ieA\cdot\dot{x}\right]\right\}
\end{equation}
where $\dot{x}_{\mu}\equiv dx_{\mu}/du$. The saddle point of the
integral over $s$ is given by
\begin{equation}
s_{0}^{2}=\frac{m^{2}}{4}\int_{0}^{1}du\,\dot{x}^{2}
\end{equation}
and its contribution to $W[A]$ is
\begin{equation}
-\sqrt{\frac{2\pi}{m}}\oint\left[dx_{\mu}\right]\frac{1}{{\textstyle \left[\int_{0}^{1}du\,\dot{x}^{2}\right]^{1/4}}}\exp\left[-m\sqrt{\int_{0}^{1}du\,\dot{x}^{2}}-ie\int_{0}^{1}du\,A\cdot\dot{x}\right].
\end{equation}
This will be a good approximation if $s_{0}\gg1$, corresponding to
$eE/m^{2}\ll1$~\cite{Dunne:2005sx}. We may now evaluate the functional
integral in steepest descents, thus reducing the problem to finding
instanton solutions for the effective action
\begin{equation}
S_{\mathit{eff}}=m\sqrt{\int_{0}^{1}du\,\dot{x}^{2}}+ie\int_{0}^{1}du\,A\left(x(u)\right)\cdot\dot{x}.\label{eq:eqm}
\end{equation}
This entails solving the following equations of motion:

\[
\frac{1}{\sqrt{\int_{0}^{1}du\,\dot{x}^{2}}}m\dot{x}_{\mu}=ieF_{\mu\nu}\dot{x}_{\nu}.
\]
Notice that $\sqrt{\int_{0}^{1}du\,\dot{x}^{2}}$ is a constant of
the motion, as can be verified by contracting the above with $\dot{x}_{\mu}.$

We impose a constant (Minkowski-space) electric field by taking $A_{3}=-iEx_{4}$
with the other three components either zero or constant. Then the
only non-zero components of $F$ are $F_{34}=-F_{43}=+iE$. The general
solution of the equations of motion~\eqref{eq:eqm} for $x_{\mu}\left(\tau\right)$
is a circular orbit of radius $R=m/eE$ centered about $(\bar{x}_{3},\bar{x}_{4})$
\begin{align}
x_{3} & =\frac{m}{eE}\cos\left(\frac{eaE}{m}u+\varphi\right)+\bar{x}_{3}\\
x_{4} & =\frac{m}{eE}\sin\left(\frac{eaE}{m}u+\varphi\right)+\bar{x}_{4}
\end{align}
where the parameter $a=\sqrt{\int_{0}^{1}du\,\dot{x}^{2}}$. In the
case of constant field this equals the arc length, which for $T=0$
is $2\pi pR$, with $p$ a positive integer. The value of the effective
action for such a solution is

\begin{equation}
S_{0p}=\frac{m^{2}}{eE}\pi p.
\end{equation}
The case $p=1$ was treated by Affleck et al., who showed that the
$p=1$ solution has one unstable mode. Their results may be extended
to general $p$ with one caveat: for $p>1$, there is an unstable
mode as in the $p=1$ case, but also $p-1$ extra pairs of negative
eigenvalues. Extra negative modes have been found previously in the
study of vacuum decay, with subtle interpretational issues~\cite{Coleman:1987rm,Liang:1994xn,Battarra:2012vu}.
In the present case, we have guidance from Schwinger's original treatment,
and this tells us that the pairs of negative eigenvalues are included,
with the minus signs cancelling. See Appendix \ref{sec:prefactors}
for a derivation of the fluctuation prefactors $K_{0p}$ associated with these trajectories.
 The final result takes the form
\begin{equation}\label{eq:Schwinger-scalar-rate-demo}
\Gamma_{0}=2\Im\left\{ \sum_{p=1}^{\infty}K_{0p}e^{-S_{0p}}\right\}
\end{equation}
with
\begin{equation}
K_{0p}=\frac{i}{2}\frac{\left(eE\right)^{2}}{\left(2\pi\right)^{3}p^{2}}\left(-1\right)^{p+1}
\end{equation}
reproducing the known result Eq. \eqref{eq:Schwinger-scalar-rate}.

\section{The $T>0$ case in the worldline formalism}

In the worldline formalism, nonzero temperature may be introduced
via the replacement~\cite{McKeon:1992if,Shovkovy:1998xw}
\begin{equation}
\left\langle x\left|e^{-s\left(-D^{2}\right)}\right|x\right\rangle \rightarrow\sum_{n\in Z}\left\langle x\left|e^{-s\left(-D^{2}\right)}\right|x+n\beta\hat{e}_{4}\right\rangle
\end{equation}
in the functional determinant. Finite temperature worldline instantons
are sections of the $T=0$ solutions whose endpoints are separated
by $n\beta$ in the time direction.

\begin{figure}[h]
\centering\input{paths.tex} \caption{\label{fig:paths}The four basic finite-temperature classical paths
in the $x_{3}-x_{4}$ plane with a winding of $\pm n$ in the timelike
direction, shown as portions of zero-temperature circular paths. All
four paths begin and end on the same two points, which are separated
by $n\beta$ in Euclidean time. From left to right, the paths are
an antiparticle long path, an antiparticle short path, a particle
short path and a particle long path. The radius $R$ is $m/eE$ and
the length of the short arcs $a_{n}$ is $R\theta_{n}$. }
\end{figure}
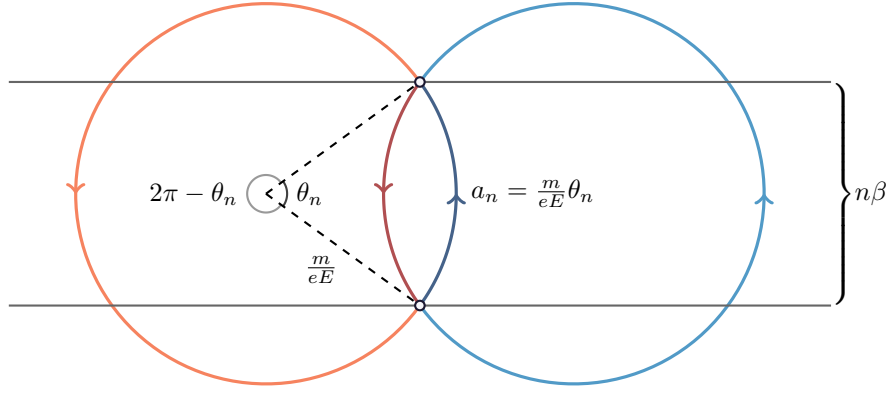

For any value of $n\beta$ for which solutions
exist, there is a short path of central angle less than $\pi$ corresponding to a particle trajectory
and another corresponding to an antiparticle trajectory; both trajectories
contribute to the free energy of the metastable phase. Correspondingly,
there are are also two long paths of central angle greater than $\pi$ which contribute to the decay rate.
All four paths are shown in Fig.~\ref{fig:paths}.
From the geometry
we see that the arc length of a short path $R\theta_{n}$ is determined by $R\sin\left(\theta_{n}/2\right)=n\beta/2$.
The arc length of a corresponding long path is $R\left(2\pi-\theta_{n}\right)$.
Appending a short
path of arc length $R\theta_{n}$ to a corresponding long path of
arc length $R\left(2\pi-\theta_{n}\right)$ gives the $T=0$ circular
solution found in Ref.~\cite{Affleck:1981bma}.

%From the geometry
%we see that $\theta_{n}$ is determined by $R\sin\left(\theta_{n}/2\right)=n\beta/2$.
%
%
%
%It is useful to distinguish between
%two types of solutions: arcs covering an angle $\theta_{n}<\mbox{\ensuremath{\pi}}$,
%which we refer to as short paths, and arcs spanning $2\pi-\theta_{n}>\pi$,
%which we call long paths.
%
%
%For any value of $n\beta$ for which solutions
%exist, there is a short path corresponding to a particle trajectory
%and another corresponding to an antiparticle trajectory; both trajectories
%contribute to the free energy of the metastable phase. Correspondingly,
%there are are also two long paths which contribute to the decay rate.
%All four paths are shown in Fig.~\ref{fig:paths}. Appending a short
%path of arc length $R\theta_{n}$ to a corresponding long path of
%arc length $R\left(2\pi-\theta_{n}\right)$ gives the $T=0$ circular
%solution found in Ref.~\cite{Affleck:1981bma}. From the geometry
%we see that $\theta_{n}$ is determined by $R\sin\left(\theta_{n}/2\right)=n\beta/2$.

%\input{filename}%\includegraphics[width=5in]{finiteTpaths2}

In order for such solutions to exist at all, the diameter of the $T=0$
solution must be greater than $n\beta$,
\begin{equation}
2R=\frac{2m}{eE}>n\beta.
\end{equation}
In other words, the maximum value of $n$, $n_{max}$ is given by
\begin{equation}
n_{max}=\left\lfloor \frac{2mT}{eE}\right\rfloor .\label{eq:nmax}
\end{equation}
This implies that there are no one-loop thermal effects from worldline
instantons for $T<eE/2m$, \emph{i.e.}, at sufficiently low temperatures~\cite{Gies:1999vb}.
%.

Such solutions can be extended by adding on $p$ windings. The actions
of these solutions are given by
\begin{align}
S_{np}^{\left(s\right)} & =\frac{m^{2}}{2eE}\left[\left(\theta_{n}+2\pi p\right)+\sin\theta_{n}\right]\\
S_{np}^{\left(l\right)} & =\frac{m^{2}}{2eE}\left[\left(2\pi-\theta_{n}+2\pi p\right)-\sin\theta_{n}\right].
\end{align}
Note that
\begin{align}
S_{np}^{\left(s\right)} & =S_{n0}^{\left(s\right)}+S_{0p}\\
S_{np}^{\left(l\right)} & =S_{n0}^{\left(l\right)}+S_{0p}
\end{align}
and that the two solutions become degenerate when $\theta_{n}=\pi.$

\subsection{Fluctuation prefactors}

The prefactors $K_{np}^{(s)}$ and $K_{np}^{(l)}$ are
given in terms of the functional determinant of the second variation
operator, which is the sum of a local and a nonlocal term~\cite{Affleck:1981bma}:
\begin{equation}
%M_{\mu\nu}=L_{\mu\nu}+c\left(x_{\mu}\left(u\right)-\bar{x}_{\mu}\right)\left(x_{\nu}\left(u'\right)-\bar{x}_{\nu}\right)$
M_{\mu\nu}\equiv\left.\frac{\delta S_{\mathit{eff}}}{\delta x_{\nu}\left(u'\right)\delta x_{\mu}\left(u\right)}\right|_{x_{\mathit{cl}}}\!=L_{\mu\nu}-\vartheta\frac{eE}{R^{2}}\left(x_{\mu}\left(u\right)-\bar{x}_{\mu}\right)\left(x_{\nu}\left(u'\right)-\bar{x}_{\nu}\right)
\end{equation}
where %\begin{equation}%L_{\mu\nu}=-\frac{eE}{\theta+2\pi p}\frac{d^{2}}{du^{2}}\delta_{\mu\nu}\delta\left(u-u'\right)+ieF_{\mu\nu}\frac{d}{du}\delta\left(u-u'\right)%\end{equation}
\begin{equation}
L_{\mu\nu}=\left[-\frac{eE}{\vartheta}\delta_{\mu\nu}\frac{d^{2}}{du^{2}}+ieF_{\mu\nu}\frac{d}{du}\right]\delta\left(u-u'\right)\label{eq:localpart}
\end{equation}
and $\vartheta$ is the total angle spanned by the instanton solution,
that is, $\vartheta=2\pi p+\theta_{n}$ for short paths and $\vartheta=2\pi\left(p+1\right)-\theta_{n}$
for long paths.
%\noindent and $c=-eE\left(2\pi p+\theta\right)/R^{2}.$
For fluctuations about zero temperature solutions, the
eigenvalue problem for $M_{\mu\nu}$ can be solved by inspection. For fluctuations about finite
temperature solutions this is made difficult by the boundary conditions at the endpoints. Fortunately, the functional
determinant may be computed without any explicit knowledge of the spectrum. The matrix
determinant lemma~\cite{Sylvester:1851fu} can be used to isolate the effect
of the nonlocal term, %\begin{equation}%\det\left[M_{\mu\nu}\right]=\det\left[L_{\mu\nu}\right]\cdot\left[1+%c\iint du\,du'\,%\left(x\left(u\right)-\bar{x}\right)_{\mu}\left(L^{-1}\right)_{\mu\nu}\left(x\left(u'\right)-\bar{x}\right)_{\nu}\right]%\end{equation}
\begin{equation}
\det\vphantom{}'\left[M_{\mu\nu}\right]=\det\vphantom{}'\left[L_{\mu\nu}\right]\cdot\left[1-\vartheta\frac{eE}{R^{2}}\iint du\,du'\,\left(x\left(u\right)-\bar{x}\right)_{\mu}\left(L^{-1}\right)_{\mu\nu}\left(x\left(u'\right)-\bar{x}\right)_{\nu}\right]
\end{equation}
and the local part of the functional determinant is computed using the method of Gel'fand and
Yaglom~\cite{Gelfand:1959nq,Levit:1976fv}.
%Pairs of negative eigenvalues
%occur for all paths with $p>0$ and are implicitly treated as in the
%$T=0$ case.
%-------------------------------------
Consider the following set
of initial value problems ($\rho=1,2,3,4$):
\begin{align}
L_{\mu\nu}\eta_{\nu}^{\left(\rho\right)} & =0\\
\eta_{\nu}^{\left(\rho\right)}\left(0\right) & =0\\
\dot{\eta}_{\nu}^{\left(\rho\right)}\left(0\right) & =\delta_{\nu\rho}
\end{align}
Up to a phase, the local prefactor can be written
\begin{equation}
\left(N\det\vphantom{}'\left[L_{\mu\nu}\right]\right)^{-1/2}=
\frac{\left(eE\right)^{2}}{\left(2\pi\vartheta\right)^{2}}\sqrt{\frac{\det\left[\tilde{\eta}_{\mu}^{\left(\nu\right)}\left(1\right)\right]}{\det\left[\eta_{\mu}^{\left(\nu\right)}\left(1\right)\right]}}
\end{equation}
where $N$ is a normalization factor and $\tilde{\eta}_{\mu}^{\left(\nu\right)}$ are the solutions of the corresponding \emph{free} initial value problem,
with $\tilde{L}_{\mu\nu}=-\frac{eE}{\vartheta}\frac{d^{2}}{du^{2}}\delta_{\mu\nu}$.
It is then straightforward to show
\begin{equation}\label{eq:localpre}
\left(N\det\vphantom{}'\left[L_{\mu\nu}\right]\right)^{-1/2}=(-1)^p\frac{\left(eE\right)^{2}}{\left(2\pi\vartheta\right)^{2}}\sqrt{\frac{\vartheta^{2}}{2\left(1-\cos\vartheta\right)}}.
\end{equation}
As we will see in section \ref{sec:topo}, the overall phase $e^{i 2\pi p /2}=(-1)^p$ is related to the Morse index of the classical path. Note that this quantity is manifestly real, which means any imaginary contribution must come from the nonlocal part.% ~\cite{morse1934calculus,Levit:1976fv}

To compute the nonlocal part we find the Green's function directly,
by solving the equation
\begin{equation}
L_{\mu\rho}G_{\rho\nu}\left(u-u'\right)=\delta_{\mu\nu}\delta\left(u-u'\right)
\end{equation}
with Dirichlet boundary conditions. A lengthy but straightforward
calculation gives, for the nontrivial components $G_{33} = G_{44}$ and $G_{43} =-G_{34}$,
\begin{align}
G_{33} &=\frac{1}{2eE}\left[-\sin\left(\vartheta|u-u'|\right)+\sin\left(\vartheta u'\right)+
\sin\left(\vartheta u\right)
-\frac{4\sin\left(\frac{\vartheta}{2}u\right)\sin\left(\frac{\vartheta}{2}u'\right)\cos\left(\frac{\vartheta}{2}\left(u-u'\right)\right)}{\tan\left(\frac{\vartheta}{2}\right)}\right]\\
G_{43} &=\frac{1}{2eE}\left[\sgn\left(u-u'\right)\left(\cos\left(\vartheta|u-u'|\right)-1\right)+\cos\left(\vartheta u'\right)-\cos\left(\vartheta u\right)-\frac{\sin\left(\vartheta\left(u-u'\right)\right)+\sin\left(\vartheta u'\right)-\sin\left(\vartheta u\right)}{\tan\left(\frac{\vartheta}{2}\right)}\right].
\end{align}

The nonlocal part of the determinant can now be computed directly:
\begin{equation}
\left[1-\vartheta\frac{eE}{R^{2}}\iint du\,du'\,x_{\mu}\left(u\right)G_{\mu\nu}\left(u-u'\right)x_{\nu}\left(u'\right)\right]=\frac{\vartheta}{2}\cot\left(\frac{\vartheta}{2}\right)
%=\begin{cases}
%\hphantom{-}\frac{1}{2}\bigl(2\pi p+\theta_{n}\bigr)\cot\bigl(\theta_{n}/2\bigr) & \text{short paths }\\
%-\frac{1}{2}\bigl(2\pi\left(p+1\right)-\theta_{n}\bigr)\cot\bigl(\theta_{n}/2\bigr) & \text{long paths}
%\end{cases}.
\end{equation}
Note that for long paths $\pi<\vartheta<2\pi$ (modulo $2\pi$), so the nonlocal part of the determinant is negative.
Because of this, it is clear that it is the long paths that contribute to the imaginary
part of the effective action.

The prefactor can now be assembled as before
\begin{align}
\left(\int d^{4}x\right)\frac{\sqrt{2\pi/m}}{\left[\int_{0}^{1}du\,\dot{x}^{2}\right]^{1/4}}\left(\det\vphantom{}'\left[\left.\frac{\delta S_{\mathit{eff}}}{\delta x_{\nu}\left(u'\right)\delta x_{\mu}\left(u\right)}\right|_{x_{\mathit{cl}}}\right]\right)^{-1/2} & =V_{3}\beta\frac{\left(eE\right)^{2}}{\left(2\pi\right)^{3/2}\left(nm\beta\right)^{1/2}\vartheta^{2}}\left[1-\left(\frac{n\beta eE}{2m}\right)^{2}\right]^{-1/4}.
\end{align}
%For long paths, a factor of $\pm i/2$ should be included.
For long paths, an extra factor of $\pm i/2$ should be included
because the contribution to the imaginary part results from an integration over only one half of the Gaussian peak in the imaginary direction~\cite{Coleman:1978ae}.
The sign depends on the way in which the analytic continuation is performed.

%-------------------------------------

%FIX

%--- COPIED FROM DISCUSSION

%---

The functional determinant for the short paths is always positive,
and the functional determinant for the long paths is always negative,
in agreement with arguments given in section~\ref{sec:topo}.
The sum of the short path contributions represents the free energy
density of the metastable phase, and is given by
\begin{equation}
f=-\sum_{p=0}^{\infty}\sum_{n=1}^{n_{\mathit{\mathit{max}}}}2K_{np}^{\left(s\right)}e^{-S_{np}^{\left(s\right)}}\label{eq:freeenergy}
\end{equation}
where
\begin{align}
K_{np}^{(s)} & =\frac{(-1)^p\left(eE\right)^{2}}{\left(2\pi\right)^{3/2}\left(n\beta m\right)^{1/2}\left[2\pi p+2\sin^{-1}\left(\frac{n\beta eE}{2m}\right)\right]^{2}}\left[1-\left(\frac{n\beta eE}{2m}\right)^{2}\right]^{-1/4}\label{eq:Ksnp}\\
S_{np}^{(s)} & =\frac{m^{2}}{2eE}\left[2\pi p+2\sin^{-1}\left(\frac{n\beta eE}{2m}\right)+\frac{n\beta eE}{m}\sqrt{1-\left(\frac{n\beta eE}{2m}\right)^{2}}\right].\label{eq:Ssnp}
\end{align}
In the limit $E\rightarrow0$, equation \eqref{eq:freeenergy} precisely
reproduces the free energy of a free relativistic particle in the
limit $\beta m\gg1$. Compare the limiting form of equation \eqref{eq:freeenergy} with the exact expression~\cite{Meisinger:2001fi}
\begin{equation}
  f = - \frac{m^2}{\pi^2 \beta^2}\sum_{n=1}^{\infty}\frac{1}{n^2}K_2(n\beta m).
\end{equation}
As $\beta \rightarrow \infty$ we have
\begin{equation}
  f \sim -  \sum_{n=1}^{\infty} 2\frac{m^{3/2}}{(2\pi)^{3/2} (n\beta)^{5/2}}e^{-n\beta m}
\end{equation}
which is the precise form of equation \eqref{eq:freeenergy} as $E\rightarrow 0$.

\begin{figure}
\centering \input{freeenergy.tex} \caption{\label{fig:freeenergy} The worldline instanton expression for the free energy is an excellent approximation at low temperatures and $E\rightarrow0$.}
\end{figure}
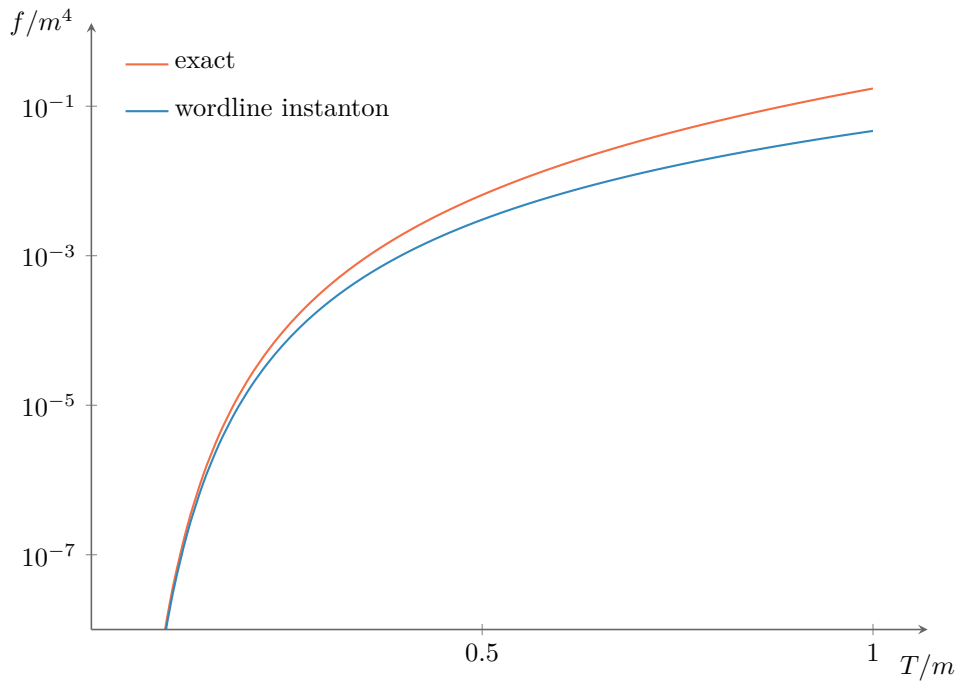

The long paths, on the other hand, give a thermal correction $\Gamma_{T}$
to the zero-temperature decay rate $\Gamma_{0}$. Our final result
for scalars is
\begin{equation}
\Gamma_{T}=2\Im\left\{ \sum_{p=0}^{\infty}\sum_{n=1}^{n_{\mathit{\mathit{max}}}}2K_{np}^{\left(l\right)}e^{-S_{np}^{\left(l\right)}}\right\} \label{eq:decayrate}
\end{equation}
where
\begin{align}
K_{np}^{\left(l\right)} & =\frac{i}{2}\cdot\frac{(-1)^p\left(eE\right)^2}{\left(2\pi\right)^{3/2}\left(n\beta m\right)^{1/2}\left[2\pi\left(p+1\right)-2\sin^{-1}\left(\frac{n\beta eE}{2m}\right)\right]^{2}}\left[1-\left(\frac{n\beta eE}{2m}\right)^{2}\right]^{-1/4}\label{eq:Klnp}\\
S_{np}^{(s)} & =\frac{m^{2}}{2eE}\left[2\pi\left(p+1\right)-2\sin^{-1}\left(\frac{n\beta eE}{2m}\right)-\frac{n\beta eE}{m}\sqrt{1-\left(\frac{n\beta eE}{2m}\right)^{2}}\right].\label{eq:Slnp}
\end{align}
This concludes the derivation of our results for scalars in
an external electric field.

\section{Discussion}

%--------------------------------------------
%
%In this section, we discuss some of the unusual features of our results,
%and their generalization to related problems. One notable feature
%is the presence of a set of thresholds controlled by the dimensionless
%parameter $2mT/eE$, and by the associated integer $n_{max}$. If
%$n_{max}=0$, \emph{i.e.} $T<eE/2m$, there are no finite temperature
%corrections to the zero-temperature decay rate. This is readly understood,
%because $2mT/eE=2R/\beta$. In order to have additional finite temperature
%contributions, we must have $2R\ge\beta$. This is not unexpected.
%In the usual treatments of the decay of metastable states of quantum
%systems at nonzero temperature REF, the zero-temperature bounce solution,
%corresponding to a critical bubble of radius $R_{c}$, is unmodified
%until $2R_{c}>\beta$. The presence of multiple thresholds, occuring
%each time $n_{max}$ increases by one as the temperature increased,
%is not as familiar. There are some similarities with the problem of
%one-loop stability of gauge fields at finite temperature in an external
%field, but the mechanism there is different, relying on the interplay
%of Matsubara frequencies with the lowest Landau level REF.
%
%--------------------------------------

%In this section, we discuss some of the unusual features of our results,
%and their generalization to related problems.

%- Intro to discussion
%- Thresholds and singularities
%- Topology (use expression with morse index)
%- results should come first

In this section, we discuss some of the unusual features of our results,
and their generalization to related problems.
In Fig.~\ref{fig:decayrate} we plot the total decay rate $\Gamma=\Gamma_{0}+\Gamma_{T}$
as a function of $T/m$ for three values of $eE/m^{2}.$ The leftmost
part of each curve represents the contribution of $\Gamma_{0}$ alone,
which is independent of temperature. Each curve shows singularities
at $T/m=neE/2m^{2}$, indicated by dotted lines. Each singularity occurs at a threshold temperature above which a new worldline instanton solution becomes possible.
It can be shown that,
wherever it is nonzero, $\Gamma_{T}$ is always larger than $\Gamma_{0}$.
By examining the reliable values of $\Gamma/m^{4}$ to the left of
each singularity, we see that the overall rise in the decay rate envelope
appears to be linear in $T$. %One-loop thermal corrections thus must
%be considered in problems where pair production in an external field
%occurs at nonzero temperature. \vspace{-0.021em}

 %\includegraphics[width=5in]{decayrate}
\begin{figure}[H]
\centering \input{decayrate.tex}

\caption{\label{fig:decayrate}The total one-loop decay rate $\Gamma=\Gamma_{0}+\Gamma_{T}$
divided by $m^{4}$ versus $T/m$ for various values of $eE/m^{2}.$
The dotted lines represent the singularities at $T/m=neE/2m^{2}$,
which are rendered finite by effects not included at one loop. The
leftmost part of each curve represents the contribution of $\Gamma_{0}$
alone. }
\end{figure}
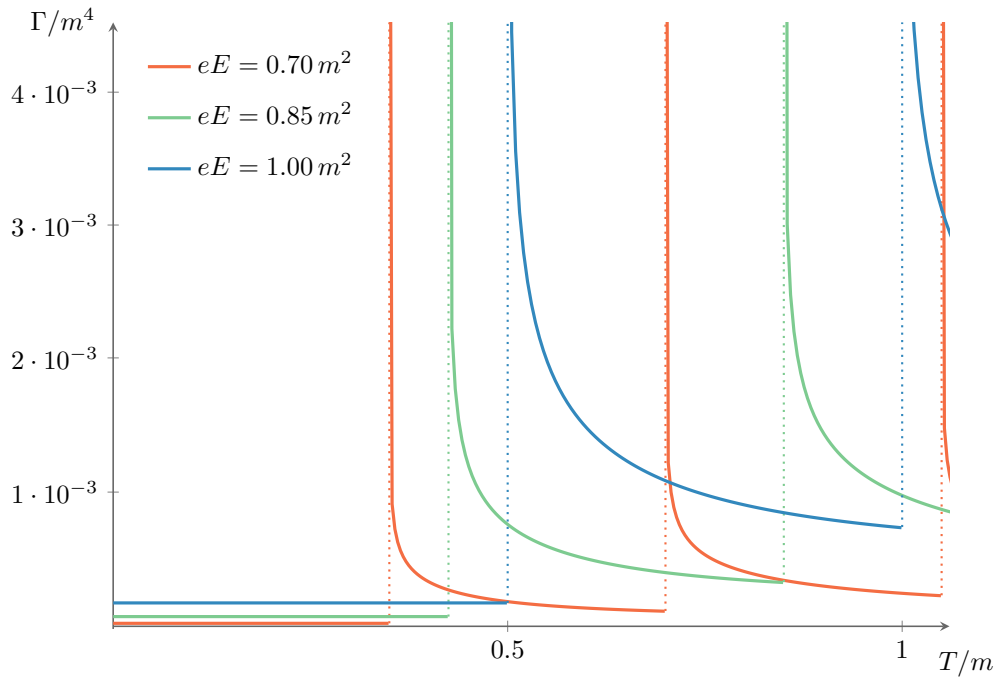

%\subsection{Thresholds and singularities}

%A notable feature of our results
%is the presence of a 
The set of thresholds is controlled by the dimensionless
parameter $2mT/eE$, and by the associated integer part $n_{max}$.
Any finite temperature instanton must satisfy $2R=2m/eE > n\beta$.
If $n_{max}=0$, \emph{i.e.} $T<eE/2m$, there are no finite temperature
instantons, and therefore no
corrections to the zero-temperature decay rate. As $T$ is increased,
the threshold for a new solution is crossed whenever $n_{max}$ increases
by one.
This has some similarity with the problem of vacuum decay at finite temperature.
In the problem of the decay of the false vacuum, the Euclidean bounce
solution in the thin wall approximation is a critical bubble of radius
$R_{c}$, obtained from the competition between volume and surface
tension contributions to the bounce action. At nonzero temperature,
this solution is unmodified until $2R_{c}>\beta$, that is, until
the bubble diameter exceeds the length of the compact direction
~\cite{Linde:1981zj,Garriga:1994ut}.
There are also some similarities with the problem of one-loop stability
of gauge fields at finite temperature in an external field, but the
mechanism there is different, and results from the competition between
positive contributions to the energy eigenvalues from Matsubara
frequencies with the negative contribution from the lowest Landau level~\cite{Meisinger:2002ji}.

\subsection{Morse-theoretic analysis}\label{sec:topo}
%\subsection{Inhomogeneous fields}\label{sec:topo}

\begin{figure}
\centering \input{action.tex} \caption{\label{fig:action}The action $S$ for an arc solution as a function
of $\theta$ with $n\beta m=1$. Only the extrema, indicated by dots,
represent classical solutions, with the short path solution a local
minimum and the long path solution a local maximum. The values of
the action at the extrema are given by $S_{n0}^{(s)}$ and $S_{n0}^{(l)}$,
respectively. At $eE/m^{2}=2$, the two extrema merge.}
\end{figure}
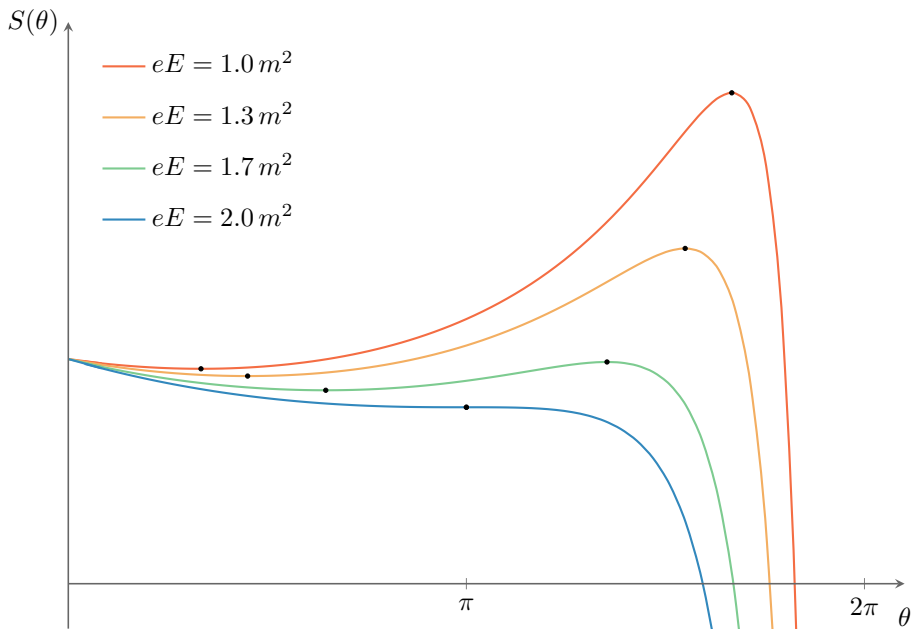

One of the striking features of our result for the decay rate,
associated with the behavior at thresholds, is the singular behavior
of the decay rate $\Gamma_{T}$. This singularity is due to the
factor
\begin{equation}\label{eq:singularity}
\left[1-\left(\frac{n\beta eE}{2m}\right)^{2}\right]^{-1/4}
\end{equation}
in eqns. \eqref{eq:Ksnp} and \eqref{eq:Klnp} for the fluctuation prefactors
$K_{np}^{(s)}$ and $K_{np}^{(l)}$. The origin of the singularity
can be understood at the classical level, however. In Figure~\ref{fig:action},
we plot $S$ as a function of the angle
$\theta$ for a family of paths with $n\beta m=1$ and fixed endpoint separation of $n\beta\hat{e}_{4}$.
Classical solutions are only obtained at the extrema, where $R=m/eE$,
and are given by $S_{n0}^{(s)}$ and $S_{n0}^{(l)}$. We see that
for $eE/m^{2}<2$, there is a local minimum corresponding to the short
path, and a local maximum corresponding to the long path. The instability
of the long path is obvious. At $eE/m^{2}=2$, the two extrema merge
and $n_{max}$ changes by one. The long and short paths are local maxima
and minima of the action, respectively, along a given direction in functional
space. Recall that $n_{max}$ is the greatest integer less than $2R/\beta=2mT/eE$.
When $2mT/eE$ is an integer, the long and short paths are degenerate,
and both are arcs of angle $\pi$ and arc length $\pi R$. If
$T$ is increased slightly, the degeneracy is lifted and a new maximum
and minimum of the action exist. The singularity in $K_{np}^{(s)}$
and $K_{np}^{(l)}$ is associated with the degeneracy of the two solutions.
This behavior is reminiscent of the behavior associated with the classical
spinodal, where a local maximum and minimum merge and the quadratic
approximation fails.

This behavior is not restricted to the case of a constant, homogeneous electric field,
but will occur generally for an inhomogeneous electric field when the temperature
is nonzero and a zero-temperature instanton exists.

Morse theory provides a useful characterization
of the eigenvalues of second variation operators about a functional extremum
~\cite{morse1934calculus,Levit:1976fv}.
The \emph{caustic} is the envelope of trajectories obtained by fixing $x_\mu(0)$
and varying $\dot{x}_\mu(0)$. A \emph{focal point} of a classical path is defined
as a contact point between the path and the caustic surface. The central point
of Morse theory may be stated thus: the number of negative eigenvalues
of the second variation operator about a given classical path equals the number
of \emph{focal points} strictly between its endpoints, where each focal point
is counted with its multiplicity. This number is the \emph{Morse index}
of the path.

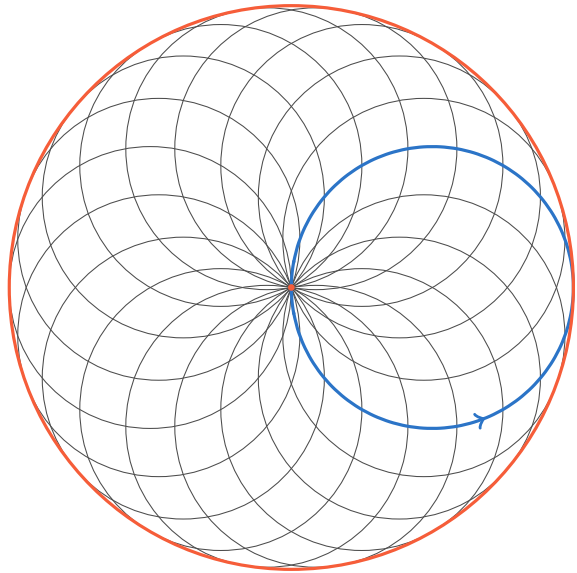
\begin{figure}
\centering \input{caustics.tex} \caption{\label{fig:caustics} A set of trajectories formed by varying $\dot{x}(0)$. The envelope of such trajectories is the caustic, indicated in red. A representative path is in blue. As the path touches the caustic, the fluctuation operator about it acquires a negative eigenvalue.}
\end{figure}

In the present case, all classical solutions are circles of constant radius $R=m/eE$.
Therefore, the \emph{caustic} is the union of a larger circle of radius $2R$ with
a single point at $x_\mu(0)$. A diagram illustrating this caustic is in figure \ref{fig:caustics}.
Zero-temperature wordline instanton solutions of winding $p$ (as well as our long
paths of the same winding) contact the caustic $2p+1$ times, which establishes their
imaginary contribution to the effective action. Short paths of winding $p$ contact the
caustic $2p$ times, which nets a real contribution to the effective action. In either case,
the $2p$ negative eigenvalues associated with additional windings result in an overall
factor of $(-1)^p$ in the prefactor, as shown in equation \eqref{eq:localpre}. The appearance
of such pairs of negative eigenvalues can be seen explicitly in the derivation of the
functional determinant prefactor for zero-temperature solutions (see Appendix \ref{sec:prefactors}).

%The envelope of classical trajectories, whenever a
%classical trajectory intersects the \emph{caustic},
%the envelope of trajectories obtained
%when $\dot{x}\left(0\right)$ is varied, the second derivative operator
%about that trajectory acquires a negative eigenvalue. The pairs of negative
%eigenvalues found in solutions with more than the
%minimum amount of windings (for zero or nonzero temperature)
%occur for the same reason: each additional winding is associated with
%two intersections with the caustic, and thus a pair of negative eigenvalues
%appears in the fluctuation spectrum. Therefore, the
%striking qualitative features of our results -- the singularities
%and thresholds, as well as the fact that short paths contribute to
%the free energy while long paths contribute an imaginary part -- are
%consequences of the topology of functional space.
%Inhomogeneities
%in the external electric field are therefore not expected to alter this
%picture significantly.

We emphasize that these features do not depend on the exact shape of the wordline instanton
solution. Exact worldline instanton trajectories are known for several inhomogeneous field
configurations. These trajectories are no longer circles, but they are still closed and
periodic~\cite{Dunne:2006st}. This is sufficient to establish that the striking qualitative
features of our results -- the singularities and thresholds, as well as the fact that short
paths contribute to the free energy while long paths contribute the decay rate -- are expected
for inhomogeneous field configurations as well.

\subsection{Higher order effects}

We have seen in previous sections that the singularities in the effective
action result from the inadequacy of the Gaussian approximation at points in
functional space where two critical points become degenerate. At such points the quadratic
coefficient in the Hessian operator vanishes and higher order terms are necessary for a correct
computation of the effective action.
The inclusion of such higher-order effects is difficult even at zero temperature.
At finite temperature, higher-order effects will give rise to finite lifetimes
for quasiparticle excitations in the thermal medium.
These lifetimes may be calculated within the hard thermal loop
(HTL) framework~\cite{Braaten:1992gd}.
It is at least plausible that these finite lifetime effects smear out
the threshold singularities, rendering the decay
rate finite.
A complete calculation of this type is far beyond our grasp.
However, the singularities can be eliminated heuristically by including the effect of a damping rate for
the charged scalars. A damping rate for scalar QED can be obtained from a hard-thermal-loop
calculation of the imaginary part of the scalar self energy.
We include this effect by replacing
$m \rightarrow m - i \gamma$
in equation \eqref{eq:singularity}, where $\gamma \sim 0.04\, e^2 T $
is the scalar damping rate~\cite{Thoma:1996ag,Abada:2005jq}. This amounts to replacing the singular factor
\begin{equation}
\left[1-\left(\frac{n\beta eE}{2m}\right)^{2}\right]^{-1/4}
\rightarrow
%\frac{\cos \left[\frac{1}{4} \tan^{-1}
%    \left[
%         \frac{2\gamma}{m}
%         \Big(\left(\frac{2 m}{n\beta eE }\right)^2-1\Big)^{-1}
%%         \left(\frac{n\beta eE }{2 m}\right)^2
%%         \Big/ \Big(1 - \left(\frac{n\beta eE }{2 m}\right)^2\left(1 - \frac{3\gamma^2}{m^2}\Big)\right)
%    \right]
%\right]}
%{
%\;\;
%\left[
%    \Big(1-\left(\frac{n\beta eE }{2 m}\right)^2\Big)^2
%    +
%    2\gamma ^2\left(\frac{n\beta eE }{2 m}\right)^2
%    \Big(3-\left(\frac{n\beta eE }{2 m}\right)^2\Big)
% \right]^{1/8}}
\left[
\frac{
1 - \left(
          \frac{n\beta eE}{2m}
    \right)^{2}
    }{
      \Big(1 - \left(
                     \frac{n\beta eE}{2m}
               \right)^{2}
      \Big)^2 + \Big(\frac{2\gamma}{m}
                     \Big(
                          \frac{n\beta eE}{2m}
                     \Big)^2
                \Big)^2
    }
\right]^{1/4}
\end{equation}
in equation \eqref{eq:Klnp}.

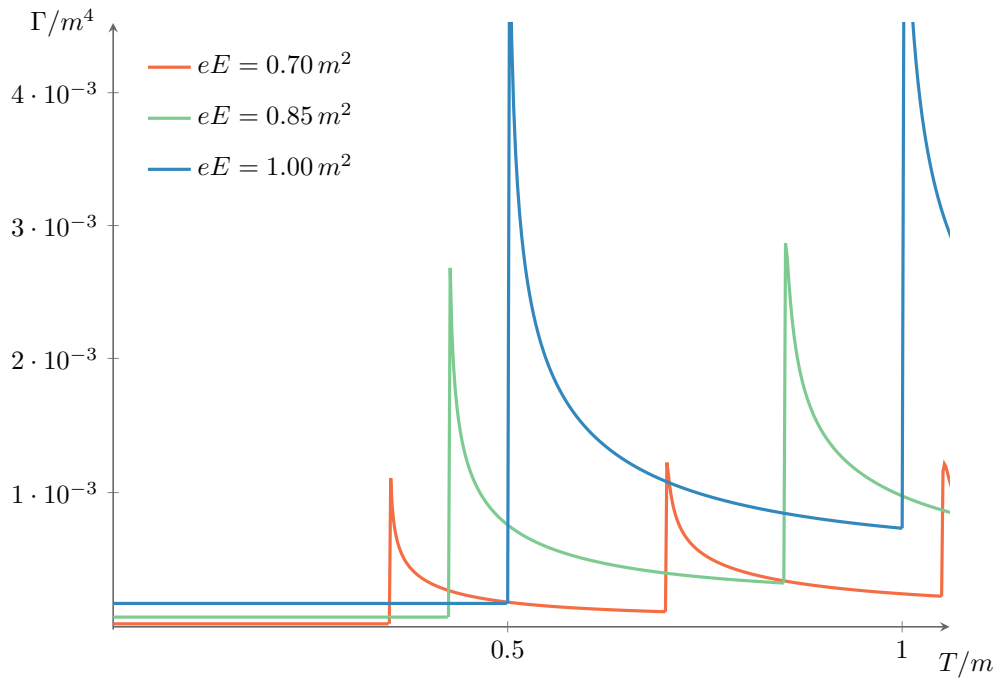
\begin{figure}[H]
\centering \input{dampeddecayrate.tex}

\caption{\label{fig:dampeddecayrate}
\label{fig:decayrate}The total one-loop decay rate $\Gamma=\Gamma_{0}+\Gamma_{T}$
divided by $m^{4}$ versus $T/m$ for various values of $eE/m^{2}$, including HTL damping effects.
The leftmost part of each curve represents the contribution of $\Gamma_{0}$
alone.
%The total one-loop decay rate $\Gamma=\Gamma_{0}+\Gamma_{T}$
%divided by $m^{4}$ versus $T/m$ for various values of $eE/m^{2}.$
%The dotted lines represent the singularities at $T/m=neE/2m^{2}$,
%which are rendered finite by effects not included at one loop. The
%leftmost part of each curve represents the contribution of $\Gamma_{0}$
%alone.
 }
\end{figure}

In Fig.~\ref{fig:dampeddecayrate} we plot the total decay rate $\Gamma=\Gamma_{0}+\Gamma_{T}$
as a function of $T/m$ with the modified threshold behavior for the same three values of $eE/m^{2}$
used in Fig.~\ref{fig:decayrate}. As in that figure, the leftmost
part of each curve represents the contribution of $\Gamma_{0}$ alone,
which is independent of temperature. Each curve shows local maxima
at $T/m=neE/2m^{2}$. There is very little difference between the unmodified
and modified decay rates, except in the close vicinity of a threshold.
We have tried other modifications of the decay rate using the HTL damping
rate. the results are not sensitive to the particular modification used,
and the decay rate away from thresholds is virtually unchanged.
This indicates that the thermal contribution to the pair production
is substantially larger than the $T=0$ contribution once the first
threshold is crossed.

\section{Conclusions}

We have presented a first-principles worldline instanton method for calculating
the thermal contributions to Schwinger pair production in an
electric field, and argued that many of the features will
carry over to the case of inhomogeneous fields as well.
While the worldline formalism is powerful, it is physically opaque.
In Appendix~\ref{sec:equivproper}, we give a formal derivation of
our results by applying a saddle-point approximation to the standard
proper time representation of the effective potential for scalar bosons
in an external electric field.
 A simple physical understanding of these results analogous to the many physically
transparent derivations of the zero temperature decay rate would also be
highly desirable.

The decay rate shows unphysical behavior at a series of
thresholds, but this is an artifact of the Gaussian approximation
to the functional integral; we have shown that this behavior
is made physical by the phenomenological inclusion of
hard thermal loop effects. Once the first threshold for thermal
effects is crossed, the thermal contribution is larger than the
the $T=0$ contribution, rising as each successive threshold is
crossed. This strongly indicates the potential importance of thermal
effects in all such nonperturbative pair production processes.
We plan to extend this work to cases of more
phenomenological interest, such as quarks in constant
non-Abelian electric fields with Polyakov loop effects
included.

After the completion of this work, a paper appeared on arXiv~\cite{Brown:2015kgj}
that considers the same problem and has some overlap with our work.
However, there is significant disagreement between our results.

%WHY DOES THIS WORK? B/C WORLDLINE FORMALISM REPLACES POLES WITH STEEPEST
%DESCENTS. WHERE DOES THIS point GO?

\appendix
\section{Fluctuation prefactor for zero temperature solution}\label{sec:prefactors}

First we illustrate the use of the matrix determinant lemma by computing
the prefactor for the zero temperature worldline solutions. Because
the local part $L_{\mu\nu}$ of the second variation operator \eqref{eq:localpart}
has zero modes associated with proper time translations and expansions/contractions
of the zero temperature circle, it is noninvertible, which means the
spectrum must be known and there is no benefit over the direct calculation.
However, at finite temperature both zero modes are lifted and the
method becomes much more convenient. The second variation of the action
about the worldline instanton solution (assumed without loss of generality
to be centered at the origin) is~\cite{Affleck:1981bma}

\begin{equation}
\left.\frac{\delta S_{\mathit{eff}}}{\delta x_{\nu}\left(u'\right)\delta x_{\mu}\left(u\right)}\right|_{x_{\mathit{cl}}}=\left[-\frac{eE}{2\pi p}\frac{d^{2}}{du^{2}}\delta_{\mu\nu}+ieF_{\mu\nu}\frac{d}{du}\right]\delta\left(u-u'\right)-2\pi p\frac{eE}{R^{2}}x_{\mu}\left(u\right)x_{\nu}\left(u'\right).
\end{equation}
The determinant can be written as

\begin{align}
\det\vphantom{}'\left[\left.\frac{\delta S_{\mathit{eff}}}{\delta x_{\nu}\left(u'\right)\delta x_{\mu}\left(u\right)}\right|_{x_{\mathit{cl}}}\right] & =\det\vphantom{}'\left[L_{\mu\nu}\right]\left(1-2\pi p\frac{ eE}{R^{2}}\iint du\,du'\,x_{\mu}\left(u\right)\left(L{}^{-1}\right)'_{\mu\nu}x_{\nu}\left(u'\right)\right)\left(-2\pi peE\right)\\
L_{\mu\nu} & =\left[-\frac{eE}{2\pi p}\frac{d^{2}}{du^{2}}\delta_{\mu\nu}+ieF_{\mu\nu}\frac{d}{du}\right].
\end{align}
As usual, the primed determinant and Green's function are computed with zero modes removed. The last factor
appears because, although changing the radius of the instanton circle is a zero
mode of the local part, it is not a zero mode of the full second
variation operator. The corresponding eigenvalue must be handled separately.

The determinant of the local part can be computed by enumerating eigenvalues,
as in Affleck et al. Ignoring the irrelevant transverse directions
we have
\begin{equation}
\det\vphantom{}'\left[L_{\mu\nu}\right]=N\prod_{\substack{q\neq0\\
q\neq p
}
}\left(2\pi eE\left(\frac{q^{2}}{p}-q\right)\right)^{2}
\end{equation}
where $N$ is a normalization factor to be fixed by the identity
\begin{equation}
\int_{x(1)=x(0)=\bar{x}}\left[dx\right]\exp\left[-\frac{m^2}{4s_{0}}\int_{0}^{1}du\,\dot{x}^{2}\right]
          =\left[N\det\vphantom{}'\left(-\frac{m^2}{4s_{0}}\frac{d^{2}}{du^{2}}\right)\right]^{-1/2}
          =\frac{m^4}{\left(4\pi s_{0}\right)^{2}}.
\end{equation}
Therefore,
\begin{align}
\det\vphantom{}'\left[L_{\mu\nu}\right] & =\left(\frac{4\pi s_{0}}{m^2}\right)^{4}\frac{1}{\left(2\pi peE\right)^{2}}\prod_{\substack{q\neq0\\
q\neq p
}
}\frac{\left(\frac{q^{2}}{p}-q\right)^{2}}{\left(\frac{q^{2}}{p}\right)^{2}}\\
 & =\left( \frac{4\pi^2 p}{eE}\right)^{4}\frac{1}{\left(2\pi peE\right)^{2}}\left[\lim_{z\rightarrow1}\frac{\sin\left(\pi pz\right)}{\pi pz\left(1-z\right)}\right]^{2}\\
 & =\left[\left(\frac{4\pi^2 p}{eE}\right)^{2}\frac{\left(-1\right)^{p+1}}{2\pi peE}\right]^{2}.
\end{align}
The second line comes from a standard identity~\cite{gradshteyn:2007aa}.

We now proceed to the nonlocal part which, as we will see, is trivial.
The Green's function $\left(L{}^{-1}\right)'_{\mu\nu}$ can be obtained
from the spectral representation
\begin{equation}
\left(L{}^{-1}\right)'_{\mu\nu}=\sum_{\substack{q\neq0\\
q\neq p
}
}\frac{1}{\lambda_{q}}\left(\begin{array}{cc}
\cos\left(2\pi q\left(u-u'\right)\right) & -\sin\left(2\pi q\left(u-u'\right)\right)\\
\sin\left(2\pi q\left(u-u'\right)\right) & \cos\left(2\pi q\left(u-u'\right)\right)
\end{array}\right)
\end{equation}
and thus
\begin{equation}
\int_{0}^{1}du\int_{0}^{1}du'\,x_{\mu}\left(u\right)\left(L{}^{-1}\right)'_{\mu\nu}x_{\nu}\left(u'\right)=\sum_{\substack{q\neq0\\
q\neq p
}
}\frac{1}{\lambda_{q}}\left(\frac{\sin\left(\pi\left(p-q\right)\right)}{\pi\left(p-q\right)}\right)^{2}=0.
\end{equation}

The last remaining piece to be evaluated is the contribution of the
zero mode associated with proper time translations $x^{\mu}\left(2\pi pu\right)\rightarrow x^{\mu}\left(2\pi pu+\varphi\right)$.
As usual, one need only consider an infinitesimal translation \foreignlanguage{english}{
\begin{equation}
x^{\mu}\left(2\pi pu+\varphi\right)\approx x^{\mu}\left(2\pi pu\right)+\varphi\left.\frac{d}{d\varphi}x^{\mu}\left(2\pi pu+\varphi\right)\right|_{\varphi=0}
\end{equation}
}and write the second term in terms of normalized eigenfunctions.
Per this standard argument, a factor of
\begin{equation}
R\int_{0}^{2\pi}\frac{d\varphi}{\sqrt{2\pi}}=\sqrt{2\pi}\frac{m}{eE}
\end{equation}
must be included in the functional integral.

Collecting everything, we obtain for the prefactor
\begin{align}
\left(\int d^{4}x\right)\frac{\sqrt{2\pi/m}}{\left[\int_{0}^{1}du\,\dot{x}^{2}\right]^{1/4}}\left(\det\vphantom{}'\left[\left.\frac{\delta S_{\mathit{eff}}}{\delta x_{\nu}\left(u'\right)\delta x_{\mu}\left(u\right)}\right|_{x_{\mathit{cl}}}\right]\right)^{-1/2}
&=\pm V_{4}\frac{i}{2}\frac{\left(eE\right)^{2}}{\left(2\pi\right)^{3}p^{2}}\left(-1\right)^{p+1}
\end{align}
in agreement with Schwinger's formula. The factor of $1/2$ comes
from integrating over only one half of the Gaussian peak in the imaginary
direction, and the sign depends on the way in which the analytic continuation
is performed.

\section{Formal equivalence with proper time formalism}\label{sec:equivproper}

The proper time expression for the one-loop finite temperature contribution
to the effective action of a charged scalar is~\cite{Gies:1998vt}
\begin{equation}
\mathcal{L}^{1T}=-\frac{(eE)^{2}}{8\pi^{2}}\sum_{n=1}^{\infty}{\int_{0}^{\infty}{\frac{ds}{s^{2}}\csc(s)e^{-\frac{m^{2}}{eE}s-\frac{eE(n\beta)^{2}}{4}\cot(s)}}}.
\end{equation}
We wish to calculate a Gaussian approximation to this integral. The
exponent has pairs of saddle points given implicitly by
\begin{equation}
\sin(s_{0})=\frac{n\beta eE}{2m}\equiv\frac{n\beta}{2R}
\end{equation}
provided the right side is smaller than $1$, that is, $n\leq n_{\mathit{max}}$
(see eq. \eqref{eq:nmax}). For definiteness and simplicity we take
$s_{0}$ to lie in the first quadrant, corresponding to our short
path solutions. The second derivative of the exponent at the saddle
point is
\begin{equation}
\left.\frac{\partial^{2}}{\partial s^{2}}\left(\frac{m^{2}}{eE}s+\frac{eE(n\beta)^{2}}{4}\cot(s)\right)\right|_{s_{0}}=\frac{eE(n\beta)^{2}}{2}\cot(s_{0})\csc^{2}(s_{0})=\frac{4m^{3}}{(eE)^{2}(n\beta)}\sqrt{1-\left(\frac{n\beta}{2R}\right)^{2}}.
\end{equation}
The Gaussian approximation to the integral reads\foreignlanguage{english}{
\begin{align}
\mathcal{L}_{(n)}^{1T} & =-\frac{\left(eE\right){}^{2}}{8\pi^{2}}\sum_{n=1}^{n_{\mathit{max}}}\frac{1}{s_{0}^{2}}\frac{2R}{n\beta}\left(2\pi\frac{(eE)^{2}n\beta}{4m^{3}\sqrt{1-(\frac{n\beta}{2R})^{2}}}\right)^{1/2}e^{-S(s_{0})}\\
 & =-\frac{(eE)^{2}}{\sqrt{2\pi^{3}}}\sum_{n=1}^{n_{\mathit{max}}}\sum_{p=0}^{\infty}\frac{1}{\left(nm\beta\right)^{1/2}\left(2\pi p+2\sin^{-1}\left(\frac{n\beta eE}{2m}\right)\right)^{2}}\left[1-\left(\frac{n\beta eE}{2m}\right)^{2}\right]^{-1/4}e^{-S(s_{0})}
\end{align}
}where
\begin{equation}
S\left(s_{0}\right)=\frac{m^{2}}{2eE}\left[2\pi p+2\sin^{-1}\left(\frac{n\beta eE}{2m}\right)+\frac{n\beta eE}{m}\sqrt{1-\left(\frac{n\beta eE}{2m}\right)^{2}}\right]
\end{equation}
in complete agreement with equations \eqref{eq:freeenergy} and \eqref{eq:Ksnp}.

%\clearpage \bibliographystyle{unsrt}
\bibliography{unLetter}

\end{document}

%% file: preamble-notikz.tex
\edef\picwidth{5.0in}
\edef\picheight{3.8in}

\usepackage{tikzexternal}

%% file: paths.tex
\tikzsetnextfilename{present-paths}
\begin{tikzpicture}[scale=1.15]

\edef\myangle{36.0}
\edef\mysmallradius{0.105}
\edef\mytinyradius{0.025}
\edef\scale{2.2}

\begin{scope}[decoration={
    markings,
    mark=at position {0.5} with {\arrow[scale=1.2]{>}}}
    ]

\draw[very thick,
      %color=sa4!85!Black
      color=sa4!85!red!85!Black,
      %color=sa4!85!red!85!,
      %color=sa4!85!red!90!Black!85,
      smooth,postaction={decorate},domain=-\myangle:\myangle] plot ({\scale*cos(\x)},{\scale*sin(\x)});
%\pgfmathparse{180-\myangle}
%\edef\startpoint{\pgfmathresult}
%\pgfmathparse{180+\myangle}
%\edef\endpoint{\pgfmathresult}
\draw[very thick,
      %color=sa2!85!Black,
      color=sa2!85!blue!85!Black,
      %color=sa2!85!blue!85!,
      %color=sa2!85!blue!90!Black!85,
      smooth,postaction={decorate},domain={180-\myangle}:{180+\myangle}] plot ( {\scale*cos(\x)+ \scale* 2 * cos(\myangle) },{\scale*sin(\x)});

\draw[very thick,
      color=sa2!85,
      %color=sa2!85!red!85,
      %smooth,
      samples=100,postaction={decorate},domain={\myangle-180}:{180-\myangle}] plot ( {-\scale*cos(\x)},{-\scale*sin(\x)});
\draw[very thick,
      color=sa4!85,
      %color=sa4!85!blue!85,
      %smooth,
      samples=100,postaction={decorate},domain={\myangle-180}:{180-\myangle}] plot ( {\scale*cos(\x)+ \scale* 2 * cos(\myangle) },{\scale*sin(\x)});

\end{scope}

\draw[thick,color=black!75,smooth,domain=-\myangle:\myangle] plot ( {1.06*\mysmallradius*\scale*cos (\x) },{1.06*\mysmallradius*\scale*sin(\x)});
%\draw[color=Gray,smooth,domain=-\startpoint:\startpoint] plot ({-.95*\mysmallradius*\scale*cos(\x)},{-.95*\mysmallradius*\scale*sin(\x)});
\draw[thick,color=Gray!80,smooth,domain={\myangle-180}:{180-\myangle}] plot ( {-0.94*\mysmallradius*\scale*cos(\x)},{-0.94*\mysmallradius*\scale*sin(\x)});

\node [right] (v1) at ( 1.06*\mysmallradius*\scale,0) {$\theta_n$};
\node [left] (v2) at ( -0.94*\mysmallradius*\scale,0) {$2\pi-\theta_n$};
\draw [thick,style=dashed,domain=0:1] plot ( {\x * \scale* cos(\myangle)}, {\x * \scale*sin(\myangle) }     );
\draw [thick,style=dashed,domain=0:1] plot ( {\x * \scale* cos(\myangle)}, {-\x * \scale*sin(\myangle)  }    );

\node (v5) at (-1.4 * \scale,{\scale * sin(\myangle)}) {};
\node (v6) at ( {\scale* (1.4 + 2* cos(\myangle) )},{\scale * sin(\myangle)}) {};
\node (v7) at ( -1.4*\scale,{-\scale * sin(\myangle)}) {};
\node (v8) at ( {\scale* (1.4 + 2*cos(\myangle))},{-\scale * sin(\myangle)}) {};

\draw [thick,black!60] (v5) edge (v6);
\draw [thick,black!60] (v7) edge (v8);

\node  at ( {\scale* (1.45 + 2* cos(\myangle) )},0) {${\LARGE\left.\begin{array}{c} \vphantom{b}\\ \vphantom{b} \\ \vphantom{b} \\ \vphantom{b} \end{array}\right\}} n\beta $};

\node [label={[label distance=-0.1*\scale cm](180+\myangle):$\frac{m}{eE}$}] (v8) at ( {\scale * cos(\myangle)/2},{-\scale * sin(\myangle)/2}) {};
\node [right] (v8) at ( {\scale*1.03 },{0}) {$ a_n=\frac{m }{eE}\theta_n$};

%\draw[color=sa5!30!Black,fill=sa5!20!Gray] ({\scale*cos(\myangle)},{\scale*sin(\myangle)}) ellipse ({\scale * \mytinyradius} and {\scale * \mytinyradius});
%\draw[color=sa5!30!Black,fill=sa5!20!Gray] ({\scale*cos(\myangle)},{-\scale*sin(\myangle)}) ellipse ({\scale * \mytinyradius} and {\scale * \mytinyradius});

\draw[thick, color=sa5!30!Black,fill=sa5!05] ({\scale*cos(\myangle)},{\scale*sin(\myangle)}) ellipse ({\scale * \mytinyradius} and {\scale * \mytinyradius});
\draw[thick, color=sa5!30!Black,fill=sa5!05] ( {\scale*cos(\myangle)},{-\scale*sin(\myangle)}) ellipse ({\scale * \mytinyradius} and {\scale * \mytinyradius});

\end{tikzpicture}

%% file: freeenergy.tex
\tikzsetnextfilename{letter-freeenergy}
\begin{tikzpicture}[scale=1.0,
    declare function={
      besselktwo(\x) = exp(-\x) * sqrt(pi / (2 * \x)) * (1 + 15/ (8 * \x) + 105/(128 * \x^2));
    }
]
    \begin{semilogyaxis}[
    log origin=infty,
      width = \picwidth,
      height = \picheight,
      xmin=0.0,
      xmax=1.07,
      ymin=0.00000001, ymax=1.3,
      axis lines=center,
      axis on top=true,
      domain=0.0:1.07,
      xtick={0,0.5,...,2},
      ytick={0.000000001,0.0000001,0.00001,0.001,0.1},
      minor y tick style={draw=none},
      legend cell align=left,
      legend image post style={xscale=0.95}, %%footnotesize
      %legend image post style={xscale=0.7},   %%small
      legend style={row sep=0.2cm,
                    fill=none,
                    /tikz/every odd column/.style={yshift=-1pt},
                    draw=none,
                    line width=0.1pt
                    },
      x label style = {at={(axis description cs:1,-0.025)}, anchor=north}, %font = \footnotesize,
      y label style = {at={(axis description cs:-0.01,1)}, anchor=east},%font = \footnotesize,
      ylabel=$f/m^4$,
      xlabel=$T/m$,
      legend pos = north west,
      mark size=0.8pt,
      %colorbrewer cycle list=Spectral,
      cycle list name=spectralmod2,
      ]

        \addplot +[smooth,samples=50,mark=none, thick, domain=0:1] { (\x^2/pi^2 * besselktwo(1/\x))  };
        \addlegendentryexpanded{$\text{exact}$}
        \pgfplotsset{cycle list shift=+2}
        \addplot +[smooth,samples=50,mark=none, thick, domain=0:1] { (\x^2/pi^2 * exp(-1/\x) * sqrt(pi / (2 / \x)))  };
        \addlegendentryexpanded{$\text{wordline instanton}$}

%        \pgfmathparse{(2 * arcsin((#1*\nbeta)/2) + sine(2 * arcsin((#1*\nbeta)/2))) / (2 * (#1*\nbeta)) }
%        \edef\myvalue{\pgfmathresult}
%        \addplot [forget plot, mark = *] coordinates {(2 * arcsin((#1*\nbeta)/2), \myvalue)};

    \end{semilogyaxis}
\end{tikzpicture} 

%% file: decayrate.tex
\tikzsetnextfilename{letter-decayrate}
\begin{tikzpicture}[scale=1.0]
    \edef\myleftedge{0}
    \edef\myrightedge{1.06}
    \edef\mybottomedge{-0.00002}
    \edef\mytopedge{0.0045}
    \begin{axis}[
      width = \picwidth,
      height = \picheight,
      xmin=\myleftedge, xmax=\myrightedge,
      ymin=\mybottomedge, ymax=\mytopedge,
      axis lines=center,
      axis on top=false,
      domain=\myleftedge:\myrightedge,
      xtick={0,0.5,...,2},
      ytick={0.001,0.002,...,0.007},
      y tick label style={
        /pgf/number format/.cd,
        %fixed,
        %fixed zerofill,
        %precision=3,
        precision=0,
        scaled y ticks=false,
        /tikz/.cd
      },
      legend cell align=left,
      legend image post style={xscale=0.95}, %%footnotesize
      %legend image post style={xscale=0.7},   %%small
      legend style={row sep=0.2cm,
                    fill=none,
                    /tikz/every odd column/.style={yshift=-1pt},
                    draw=none,
                    line width=0.1pt
                    },
      %x label style = {at={(axis description cs:1,-0.06)}, anchor=north},%font = \footnotesize,
      x label style = {at={(axis description cs:1.02,-0.02)}, anchor=north},%font = \footnotesize,
      y label style = {at={(axis description cs:-0.01,1)}, anchor=east},%font = \footnotesize,
      ylabel=$\Gamma / m^4$,
      xlabel=$\displaystyle T/m$,
      restrict x to domain=0:2,
      restrict y to domain=-0.1:0.1,
      legend pos = north west,
      mark size=0.8pt,
      %colorbrewer cycle list=Spectral,
      %colorbrewer cycle list=YlGnBu,
      %cycle list name=exotic,
      cycle list name=spectralmod,
      ]
      \pgfplotsset{cycle list shift=+1}

      \foreach \eEoverMsq in {0.70,0.85,1.00}{%
      %\foreach \eEoverMsq in {0.70,0.80, 0.90, 1.00}{%
      %\foreach \eEoverMsq in {0.60,0.80,1.00}{%
      %\foreach \eEoverMsq in {1.00,0.85,0.70}{%
          \addplot+[forget plot, mark = none, very thick] gnuplot[raw gnuplot] {%
            %set samples (\spike > 0) ? 30 : 100;
            set samples 100;
            \GammaTtermDefine
            \DecayRateDefine
            myqperiod = \eEoverMsq / 2.0;
            numspikes = (ceil((\myrightedge - \myleftedge) / myqperiod) - 1);
            plot[t=0.00:myqperiod - 0.001] Decayrate(\eEoverMsq,t);
            do for [i=1:numspikes] {
              %plot[t=i*myqperiod + 0.0009:(i+1)*myqperiod - 0.0009] Decayrate(\eEoverMsq,t);
              plot[t=i*myqperiod + 0.000001:(i+1)*myqperiod - 0.0009] Decayrate(\eEoverMsq,t);
             % plot[t=i*myqperiod + 0.0001:(i+1)*myqperiod - 0.0009] Decayrate(\eEoverMsq,t);
            }
          };
          \addplot+[forget plot, mark = none,
          style=dotted,
          thick] gnuplot[raw gnuplot] {%
            %set samples (\spike > 0) ? 30 : 80;
            \GammaTtermDefine
            \DecayRateDefine
            myqperiod = \eEoverMsq / 2.0;
            numspikes = (ceil((\myrightedge - \myleftedge) / myqperiod) - 1);
            set parametric;
            do for [i=1:numspikes] {
              plot [t=Decayrate(\eEoverMsq, i * myqperiod-0.01):\mytopedge] i * myqperiod, t;
            }
          };

         \addplot+ [very thick, domain=0.001:0.005](0.51,-x);
         %\addlegendentryexpanded{$\tfrac{eE}{m^2} = \number\eEoverMsq$}
         \addlegendentryexpanded{$eE = \number\eEoverMsq\, m^2$}
      }
    \end{axis}
  \end{tikzpicture} 

%% file: action.tex
\tikzsetnextfilename{letter-action}
\begin{tikzpicture}[scale=1.0,
    declare function={
      arcsin(\x) = rad(asin(\x));
      sine(\x) = sin(deg(\x));
      Snormalized(\x,\y,\z) = \x / (2 * sine (\x / 2)) - \y * \z * (\x - sine (\x))/(8 * (sine (\x/2))^2);
    }]
    \begin{axis}[
      width = \picwidth,
      height = \picheight,
      %width = 0.55\textwidth,
      %height = 0.45\textwidth,
      xmin=0, %xmax=6.28319,
      xmax=6.6,
      ymin=-0.2, ymax=2.5,
      axis lines=center,
      axis on top=true,
      domain=0.001:6.28318,
      xtick={0,3.14159,6.28318},
      xticklabels={$0$,$\pi$,$2\pi$},
      ytick=\empty,
      legend cell align=left,
      legend image post style={xscale=0.95}, %%footnotesize
      %legend image post style={xscale=0.7},   %%small
      legend style={row sep=0.2cm,
                    fill=none,
                    /tikz/every odd column/.style={yshift=-1pt},
                    draw=none,
                    line width=0.1pt
                    },
      x label style = {at={(axis description cs:1,0.05)}, anchor=north}, %font = \footnotesize,
      y label style = {at={(axis description cs:0.0,1)}, anchor=east},%font = \footnotesize,
      ylabel=$S(\theta)$,
      xlabel=$\theta$,
      restrict y to domain=-10:10,
      legend pos = north west,
      mark size=0.8pt,
      %colorbrewer cycle list=Spectral,
      cycle list name=spectralmod2,
      ]
      \edef\nbeta{1.0}
      \pgfplotsinvokeforeach{1.0,1.3,1.7,2.0} {
      %\pgfplotsinvokeforeach{0.2,0.3,0.4,0.5} {
      %\pgfplotsinvokeforeach{0.45,0.60,0.75,1.00} {
        \addplot +[forget plot,smooth,mark=none, thick, restrict x to domain=0.00:0.28] {1-(#1)*\nbeta*\x / 12 + \x^2 / 24};
        \addplot +[smooth,samples=50,mark=none, thick, restrict x to domain=0.002:6.28318] {Snormalized(x,\nbeta,(#1))};

        \pgfmathparse{(2 * arcsin((#1*\nbeta)/2) + sine(2 * arcsin((#1*\nbeta)/2))) / (2 * (#1*\nbeta)) }
        \edef\myvalue{\pgfmathresult}
        \addplot [forget plot, mark = *] coordinates {(2 * arcsin((#1*\nbeta)/2), \myvalue)};
        \pgfmathparse{-((#1*\nbeta) * sqrt(4 - (#1*\nbeta)^2) - 4 * pi + 4 * arcsin((#1*\nbeta)/2))/(4 * (#1*\nbeta)}
        \edef\myvalue2{\pgfmathresult}
        \addplot [forget plot, mark = *] coordinates {( 2 * pi - 2 *  arcsin(#1*\nbeta/2), \myvalue2)};

        \addlegendentry{$eE = #1\,m^2$}
      }
    \end{axis}
  \end{tikzpicture} 

%% file: caustics.tex
\tikzsetnextfilename{letter-caustics}
\begin{tikzpicture}[scale=0.75]
\pgfmathsetmacro{\scale}{2.5}
\pgfmathsetmacro{\myangle}{20}

\foreach \n in {1,...,17} {%{1,...,17} {
  \pgfmathsetmacro{\infloop}{0+\myangle*\n}
  \pgfmathsetmacro{\suploop}{360+\myangle*\n}
  \pgfmathsetmacro{\centerx}{\scale*cos(\n*\myangle)}
  \pgfmathsetmacro{\centery}{\scale*sin(\n*\myangle)}
  \draw[ %thick,
  color=Gray!60!Black,smooth,%samples=80,
  %samples=10,
  %domain=\infloop:\suploop
  ] %(\scale*cos(\n*\myangle),\scale*sin(\n*\myangle)) ellipse (0.5*\scale and 0.5*\scale);
  (\centerx,\centery) circle (\scale);
  %plot ({\scale*cos(\x)+\scale*cos(\n*\myangle)},{\scale*sin(\x)+\scale*sin(\n*\myangle)});
}

%\draw[very thick,color=sa2!85!red,smooth cycle,samples=100,tension=0.1,
%      domain=180:540] plot ({2*\scale*cos(\x)},{2*\scale*sin(\x)});

\draw[very thick,color=sa4!85!blue,
       %domain=180:540,
       domain=190:530,
       samples=50,
       smooth ,%samples=10,
       decoration={
         markings,
         mark=at position {0.30} with {\arrow[scale=1.1]{>}},
       },postaction={decorate}
] plot ({\scale*cos(\x)+\scale*cos(0*\myangle)},{\scale*sin(\x)+\scale*sin(0*\myangle)});
\draw[very thick,color=sa4!85!blue,
       %domain=180:540,
       domain=-11:11,
       samples=50,
       smooth ,%samples=10,
] plot ({-\scale*cos(\x)+\scale*cos(0*\myangle)},{-\scale*sin(\x)+\scale*sin(0*\myangle)});

%\draw[very thick,color=sa4!85!blue,
%       domain=180:540,
%       samples=50,
%       smooth cycle ,%samples=10,
%       decoration={
%         markings,
%         mark=at position {0.293} with {\arrow[scale=1.1]{>}},
%       },postaction={decorate}
%] plot ({\scale*cos(\x)+\scale*cos(0*\myangle)},{\scale*sin(\x)+\scale*sin(0*\myangle)});

\draw[very thick,color=sa2!85!red] circle (2*\scale);

%\draw [very thick,color=sa4!85!blue] circle (0.03);
\draw [very thick,color=sa2!85!red] circle (0.03);
\end{tikzpicture} 

%% file: dampeddecayrate.tex
\tikzsetnextfilename{letter-dampeddecayrate}
\begin{tikzpicture}[scale=1.0]
    \edef\myleftedge{0}
    \edef\myrightedge{1.06}
    \edef\mybottomedge{-0.00002}
    \edef\mytopedge{0.0045}
    \begin{axis}[
      width = \picwidth,
      height = \picheight,
      xmin=\myleftedge, xmax=\myrightedge,
      ymin=\mybottomedge, ymax=\mytopedge,
      axis lines=center,
      axis on top=false,
      domain=\myleftedge:\myrightedge,
      xtick={0,0.5,...,2},
      ytick={0.001,0.002,...,0.007},
      y tick label style={
        /pgf/number format/.cd,
        %fixed,
        %fixed zerofill,
        %precision=3,
        precision=0,
        scaled y ticks=false,
        /tikz/.cd
      },
      legend cell align=left,
      legend image post style={xscale=0.95}, %%footnotesize
      %legend image post style={xscale=0.7},   %%small
      legend style={row sep=0.2cm,
                    fill=none,
                    /tikz/every odd column/.style={yshift=-1pt},
                    draw=none,
                    line width=0.1pt
                    },
      %x label style = {at={(axis description cs:1,-0.06)}, anchor=north},%font = \footnotesize,
      x label style = {at={(axis description cs:1.02,-0.02)}, anchor=north},%font = \footnotesize,
      y label style = {at={(axis description cs:-0.01,1)}, anchor=east},%font = \footnotesize,
      ylabel=$\Gamma / m^4$,
      xlabel=$\displaystyle T/m$,
      restrict x to domain=0:2,
      restrict y to domain=-0.1:0.1,
      legend pos = north west,
      mark size=0.8pt,
      %colorbrewer cycle list=Spectral,
      %colorbrewer cycle list=YlGnBu,
      %cycle list name=exotic,
      cycle list name=spectralmod,
      ]
      \pgfplotsset{cycle list shift=+1}

      \foreach \eEoverMsq in {0.70,0.85,1.00}{%
      %\foreach \eEoverMsq in {0.70,0.80, 0.90, 1.00}{%
      %\foreach \eEoverMsq in {0.60,0.80,1.00}{%
      %\foreach \eEoverMsq in {1.00,0.85,0.70}{%
          \addplot+[forget plot, mark = none, very thick] gnuplot[raw gnuplot] {%
            %set samples (\spike > 0) ? 30 : 100;
            set samples 200;
            \DampingRateDefine
            \DampedGammaTtermDefine
            \DampedDecayRateDefine
            myqperiod = \eEoverMsq / 2.0;
            numspikes = (ceil((\myrightedge - \myleftedge) / myqperiod) - 1);
            plot[t=0.00:myqperiod - 0.001] DampedDecayrate(\eEoverMsq,t);
            do for [i=1:numspikes] {
              %plot[t=i*myqperiod + 0.0009:(i+1)*myqperiod - 0.0009] DampedDecayrate(\eEoverMsq,t);
              plot[t=i*myqperiod + 0.000000000000000001:(i+1)*myqperiod - 0.00000000000000001] DampedDecayrate(\eEoverMsq,t);
             % plot[t=i*myqperiod + 0.0001:(i+1)*myqperiod - 0.0009] DampedDecayrate(\eEoverMsq,t);
            }
          };
          \addplot+[forget plot, mark = none,
          style=dotted,
          thick] gnuplot[raw gnuplot] {%
            %set samples (\spike > 0) ? 30 : 80;
            \DampingRateDefine
            \DampedGammaTtermDefine
            \DampedDecayRateDefine
            myqperiod = \eEoverMsq / 2.0;
            numspikes = (ceil((\myrightedge - \myleftedge) / myqperiod) - 1);
            set parametric;
            do for [i=1:numspikes] {
              %plot [t=DampedDecayrate(\eEoverMsq, i * myqperiod-0.01):DampedDecayrate(\eEoverMsq, i * myqperiod+0.000001)] i * myqperiod, t;
            }
          };

         \addplot+ [very thick, domain=0.001:0.005](0.51,-x);
         %\addlegendentryexpanded{$\tfrac{eE}{m^2} = \number\eEoverMsq$}
         \addlegendentryexpanded{$eE = \number\eEoverMsq\, m^2$}
         
      }

    \end{axis}
  \end{tikzpicture} 